\documentstyle[editedvolume]{crckapb} 

\begin{opening}
\title{GR15 session B.3 -- Physics of the Early Universe}

\author{Jo\~ao Magueijo}
\institute{Imperial College, Theoretical Physics,\\
           Prince Consort Road, London SW7 2AZ, United Kingdom}

\end{opening}

\runningtitle{GR15 session B.3 -- Physics of the Early Universe}

\begin{document}


\section{Introduction}
Session B.3 received a partisan organisation, and was divided into 
sections corresponding to the main paradigms pervading modern 
cosmology. Three sub-sessions were allocated to cover
inflationary cosmology, pre-big-bang scenarios, and topological
defects in cosmology. Anything not fitting into these topics makes up the
last section, covering miscellaneous topics.

Below I briefly review the current status in each subject
covered in a sub-session after which I summarise the talks presented. 
These summaries reflect my personal understanding of the talks, and 
I apologise to the speakers if I accidentally missed the entire point.

\section{Pre Big Bang cosmology}
The pre-big-bang scenario was proposed by Veneziano in 1991 
and Gasperini and Veneziano in 1993 and
became the subject of much interest in the past year. It is inspired by
string theory, and uses as a starting point the four-dimensional
low energy effective action of string theory. Cosmological solutions
to this action exhibit two branches. One starts in the weak coupling regime
and is driven towards strong coupling and (super-)inflationary expansion.
The other corresponds to the standard Big Bang model, with decelerated
expansion. Strong coupling effects in string theory are usually invoked 
to justify the jump from one branch to the other. 
The ensuing scenario combines a 
super-inflationary branch, which solves the usual cosmological
problems solved by inflation, with a ``graceful exit'' into a 
post-inflationary branch. Clearly a large number of details remain
to be filled in. However attention in the past year has concentrated on
how much fine tuning the model requires, regardless of its uncertainties.
In particular there have been claims that the model requires fine
tuning in the initial conditions, eg: already a very flat Universe at 
the start,
or homogeneity across a large number of Planck size regions.

Two very interesting talks were given in this session. Dominic Clancy
presented work in which Pre Big Bang scenarios were reexamined within
the broader class of anisotropic cosmological models. The pre big bang
model was initially proposed using a flat Friedmann  model as its setting. 
The model appeared to be considerably more contrived when the effects
of curvature were considered. In the work presented in this talk
it was found that the combined effects of curvature and anisotropy would
tighten further the conditions for a successful model. It was found that in
this class of models the onset of inflation may even be prevented. In general 
the conditions for a successful model appear to be more stringent than 
for isotropic models.

David Wands gave a talk on the perturbation spectrum in these scenarios.
The graviton and dilaton perturbations produced on large scales 
were found to have steep blue spectra that leave essentially 
no perturbations on large scales. However other massless fields, 
including antisymmetric tensor fields also present in
the low-energy string action, may have very different perturbation
spectra. It was shown how the symmetries of the action can be
used to calculate the perturbations spectra. In some cases (eg.
axion fields) the perturbations on large-scales may have a 
scale-invariant Harrison-Zel'dovich spectrum suitable for producing
large-scale structure in our universe.

\section{Inflationary cosmology}
Inflationary cosmology has become the mainstream of modern cosmology.
It has become so trendy that quite frankly I can't be bothered to
review it, and I will proceed to summarise the talks given in the
corresponding session. 

A very interesting variation on inflationary scenarios, labelled warm
inflation, was presented by Arjun Berera. In this scenario, the standard
supercooling followed by reheating is bypassed. Instead, 
the radiation energy density smoothly decreases
during an inflationlike stage and with no discontinuity enters the
subsequent radiation-dominated stage. Scale factor solutions in this
scenario which do not conflict with observations  were presented.
It was shown that these scenarios predict a unique feature,
which should distinguish them from standard inflation: a large angle
cut off in the cosmic radiation anisotropy. Likelihood fits to DMR
4 year maps were presented. Finally some discussion was given of
thermally induced density fluctuations in these models. 

Reheating in standard inflation was the subject of two talks.
In the first Bruce Basset discussed geometric reheating after 
Inflation. It was shown that purely gravitational couplings during 
reheating induce parametric production of scalar, vector and tensor 
particles from the vacuum due to the coherent energy 
density oscillations of the inflaton. Implications to baryogenesis
were discussed. Furthermore the resonant amplification of the 
tensor perturbations was found to be wavelength dependent, thus breaking
the scale invariance of the gravitational wave spectrum.

In the second talk on reheating, Stephan Ramsey talked on the 
non equilibrium inflaton 
dynamics and reheating. An analysis of nonperturbative,
nonequilibrium dynamics of a quantum field in the reheating phase of
inflationary cosmology was presented. This included the full back 
reaction of the quantum field on the curved spacetime, as well as 
the fluctuations on the mean field. Several corrections and criticisms
to the standard lore were presented.

Open inflation is another popular area of research, and Takahiro Tanaka
gave a talk on the initial spectrum of cosmological perturbations
in the one bubble open Universe. The dynamics of the quantum fluctuations 
of the  inflaton field with $\lambda\phi^4$ potential was studied. 
Several results were presented aiming at verifying the ansatz of the 
quantum-to-classical transition that is often assumed in the standard 
evaluation of the amplitude of the primordial fluctuations.

Finally Valerio Faraoni examined the effects on inflation
of non--minimal coupling of the scalar field. It was argued that
the coupling $\xi$ of the scalar field to the Ricci curvature in a
curved spacetime is not a free parameter. From this point of view
the consistency of the most popular inflationary scenarios was evaluated.
Observational constraints on $\xi$ were discussed.

\section{Topological defects in cosmology}
The last year saw a number of papers dismissing topological
defects as candidates for structure formation. Ruling out
defects even became fashionable among certain levels of society. Improved
calculations on more powerful computers are normally blamed
for this. In some cases it also now appears that defects' inability
to generate large scale structure is a generic feature of scaling
defects. The fact that the population of scientists working in the field
has by no means decreased testifies to how seriously these claims
are being taken. On the other hand it is true that concrete rebuffal of these
claims, based on quantitative work of  comparable quality, has been 
slow to emerge. 

In any case  there is clearly room for plenty of work on
the field.
In particular it is clear that the uncertainties 
plaguing calculations, most notably for cosmic strings, are still there 
as much as they were before. A better assessment of the systematic errors
in all the simulations is still required. In fact in the case of local 
cosmic strings a proper computer simulation is still missing.
As for the arguments concerning generic defects, it is clear that
the parameter space surveyed is rather limited. In fact it is not even clear
that it covers local cosmic strings. 

A most interesting talk was given by Supratim Sengupta, in which
a new mechanism for defect-antidefect pair production was discussed.
The mechanism was illustrated by numerical work on first order phase 
transitions in a global U(1) theory, but the results may be applicable
in more general theories, and in second order phase transitions. 
It was shown that as the true vacuum bubbles grow and collide, the kinetic
energy released may flip the phase of the order-parameter field, 
resulting in the formation of a vortex-antivortex pair. Conditions
for this process to take place were identified, and its efficiency
compared with the Kibble mechanism. It was found that if the nucleation
rate is low, and damping not too important, this mechanism may dominate
over the Kibble mechanism. If this is the case, no doubts the standard picture
of defect formation will require extensive revision. More interesting still
is the possibility that a different scaling solution will be achieved if,
as is the case with this mechanism, only string loops are formed during the
phase transition.

Brandon Carter then presented some work on string loop dynamics and 
vorton formation. Recent progress in the treatment of the dynamics 
of conducting cosmic strings was reviewed, with emphasis on 
the allowance for electromagnetic self interaction. This formalism
was then applied to the problem of evolution of circular loops
towards ``vorton'' equilibrium states.

There were two talks in which topological defects and inflation appeared
combined together. Nobuyuki Sakai discussed a particular brand of 
inflationary scenarios in which inflation starts inside a topological defect,
most notably Planck scale monopoles. In this talk, this so-called 
topological inflation, was reanalysed in the context of
Brans-Dicke gravity. It was found that any expanding monopole 
eventually turns to contract and becomes stable, contrary to the 
case in the Einstein theory. It was then stressed that as a realistic
model, however, this model could not avoid fine-tuning of the coupling 
constant. Shinta Kasuya, on the other hand, discussed the formation of 
topological defects during preheating. The dynamics of a scalar field 
$\Phi$ with potential $g(|\Phi|^2-\eta^2)^2/2$ was studied, and 
fluctuations due to different effects were examined: namely fluctuations due
to parametric resonance and the negative curvature of the potential.
It was found that over a large parameter region GUT defects will not be
produced after inflation.

Finally Anjan Ananda Sen reported work on the gravitational fields of 
both local and global cosmic strings in the context
of Brans-Dicke theory. A class of solutions for global strings
was presented, and inconsistencies plaguing the local string case were 
pointed out. The motion of photons and material particles in these 
space-times was also examined.

\section{Miscellaneous}
The session closed with four talks in which various general problems in
cosmology were addressed. Daksh Lohiya reexamined nucleosynthesis in
his alternative cosmological scenario. Neeraj Upadhyaya addressed the
baryogenesis problem from the point of view of primordial black holes.
Amitabh Mukerjee presented a study of a phase transition. Yasusada Nambu
talked on cosmological perturbation theory.

In Daksh Lohiya's model of the Universe the scale factor $a$ is proportional
to the cosmic time $t$. The flatness, horizon, and cosmological constant
problems appear to be avoided. However in this talk the focus was on 
how nucleosynthesis proceeds in this scenario. 
It was found that right amounts of 
Helium and other heavy elements are obtained. On the other hand, deuterium, 
and other light elements, are not produced in sufficient amounts. 

Neeraj Upadhyaya gave a talk on black hole baryogenesis. The 
evolution of the masses of a collection of black holes created 
immediately at the end of inflation was studied, taking into 
account both the accretion of background matter by the black holes 
as well as the mass loss due to Hawking emission.
The baryon excess produced in this scenario was evaluated and
the implications for the current matter-antimatter asymmetry discussed.

Amitabh Mukerjee gave a very interesting report on work under way
aiming to study a phase transition in $\varphi^4$ theory
with $\varphi^3$ symmetry breaking. This numerical work is
motivated by recent work on the one-loop effective potential in
$\varphi^4$ field theory at finite temperature. The main purpose of this
work is to determine the nature of the phase transition. 

The session closed with a talk by Yasusada Nambu on the solution 
of the gauge invariant cosmological perturbation equations in 
long-wavelength limit. A new method for deriving exact solutions
of the gauge invariant perturbation equations in a flat FRW
universe with scalar fields was presented. This method finds
pertinent application in the case where there are many scalar
fields.

\section*{Acknowledgements} I would like to thank the Local Organising
Committee for the large amounts of effort put into the organisation of 
a very successful conference. Warm thanks to  Garima, Neelan, and 
Gaurav Gupta for ensuring my mental sanity on alien ground. Finally 
I thank Kim Baskerville for reading the manuscript, and
the Royal Society for financial support. 

\end{document}